\documentclass[preprint,12pt,authoryear]{elsarticle}
\usepackage[utf8]{inputenc}
\usepackage{amsmath}
\usepackage{graphicx}
\usepackage[section]{placeins} 
\usepackage{float}
\journal{Mechanics of Materials}
\begin{document}

\begin{frontmatter}

\title{Examining the Impact of Asymmetry in Lattice-Based Mechanical Metamaterials}       
\author[inst1]{Srikar Srivatsa}

\affiliation[inst1]{organization={Sibley School of Mechanical and Aerospace Engineering, Cornell University},
addressline={130 Upson Hall}, 
city={Ithaca},
postcode={14853}, 
state={New York},
country={USA}}

\author[inst2]{Roshan Suresh Kumar}

\affiliation[inst2]{organization={Department of Aerospace Engineering, Texas A\&M University},
addressline={701 H.R. Bright Bldg}, 
city={College Station},
postcode={77843}, 
state={Texas},
country={USA}}

\author[inst2]{Daniel Selva}
\author[inst1]{Meredith N. Silberstein}

\begin{abstract}
Lattice-based mechanical metamaterials can be tailored for a wide variety of applications by modifying the underlying mesostructure.  However, most existing lattice patterns take symmetry as a starting point.  We show that asymmetric lattice patterns can be more likely to have certain mechanical properties than symmetric lattice patterns.  To directly compare the effects of asymmetric versus symmetric lattice arrangements, a constrained design space is defined.  A generative design process is used to generate both symmetric and asymmetric lattice patterns within the design space.  Asymmetric lattice patterns are shown to have distinct metamaterial property spaces from symmetric lattice patterns.  Key design features are identified that are present predominantly in asymmetric lattice patterns.  We show that asymmetric lattice patterns with two of these features (\textit{arrows} and \textit{spider nodes}) are more likely to induce a broader range of Poisson’s ratios and larger shear stiffness values, respectively, compared to lattice patterns without these features.  In addition, we show that symmetry can play a role in hampering the impact of multiple features when present.  This work provides insights into the benefits of using asymmetric lattice patterns in select metamaterial design applications.
\end{abstract}

\end{frontmatter}

\section{Introduction}
\label{sec:introduction}

Mechanical metamaterials have mechanical properties governed by material mesostructure in addition to the intrinsic material properties \citep{Surjadi2019, Mir2014, Christensen2015, Lee2012}. This ``properties by mesostructure'' concept is widespread in natural materials such as sponges, bone, and bamboo \citep{Gibson1997,Gibson2005,Wegst2015,Habibi2014,Habibi2015,Libonati2017}.  Synthetically-produced mechanical metamaterials have deliberately-arranged structures that can be tailored for desirable mechanical characteristics such as high stiffness-to-density ratios, negative Poisson's Ratios, or tunable vibration control \citep{Yu2018,Chen2017,Kadic2012,Vogiatzis2017}. The most significant catalyst for mechanical metamaterial development in recent years has been advances in additive manufacturing.  These advances mean additive manufacturing can accommodate features as small as nanometers, and utilize a wide range of constituent materials, enabling the creation of structural metamaterials with a variety of mesostructure patterns \citep{Raney2015,Truby2016,Yang2019,Berger2017,Wang2018,Morita2021}.

Lattices are one of the most prominent mechanical metamaterial patterns.  Metamaterials with lattice patterns offer a large degree of customizability within a single unit cell \citep{Yu2018,Adhikari2021,Jia2020}.  The building blocks of most lattice structures are slender members that connect to each other at junctions.  The properties of a metamaterial with a lattice structure are set by varying the number, size, shape, and connection points of these members.  Plate- and shell-lattice patterns, which use a similar basic layout to member-based lattices, are less common since their larger local geometries offer fewer modes of customizability than member-based lattice patterns \citep{Queheillalt2005,Wang2018,Bonatti2019b,Evans2015}.  The versatility of member-based lattice metamaterials is prominent in design optimization studies; the relatively low volumes of individual members allow for sufficient space to optimize the member geometries and locations\citep{Asadpoure2015,Wang2018}.  The high degree of flexibility in the arrangements of member-based lattices means that lattice metamaterials allow for a broad design space even if only a few design variables are used \citep{Abdeljaber2016,Abdelhamid2018}.  This flexibility means that lattice patterns can easily be used to realize graded properties \citep{Turco2017,Jenett2020,Tancogne-Dejean2018}.  Additionally, the modularity of member-based lattice patterns can allow unit cells with different patterns to be assembled together, making them widely useful and even capable of forming hierarchical structures \citep{Vangelatos2019,Mizzi2020,Kaur2017}.

In terms of unit cell geometry, broadly speaking, metamaterial designs are either symmetric or asymmetric.  Symmetric lattice patterns are often based on existing, well-known arrangements of lattice members that offer predictable stiffness-to-density ratios.  Furthermore, many of these patterns have been studied with a variety of constituent materials, aspect ratios, and in combination with other lattice patterns \citep{Turco2017,Mizzi2020,Abdelhamid2018,Vangelatos2020}.  As such, the behavior of symmetric lattices is thoroughly documented in existing metamaterial design work.  However, many symmetric lattice patterns are limited to offering cubic characteristics \citep{Berger2017,Asadpoure2015,Spadoni2012,Wang2020a}.  Conversely, asymmetric lattice patterns can be useful in achieving anisotropic properties.  Many of the asymmetric lattice patterns described in literature are limited to distortions of existing symmetric patterns.  These distortions are most often manifested as changes to the thickness or lengths of specific members, members with variable cross-sections, curved members, or a distortion of the entire unit cell \citep{Tancogne-Dejean2018,Portela2020,Bonatti2019,Wang2021}. Interestingly, some of these distorted patterns have been shown to display quasi-isotropic properties \citep{Horrigan2009,Mizzi2015}.  Additionally, many chiral lattice patterns (that are by definition asymmetric) result in some form of isotropy \citep{Gatt2013,Carta2016,Mizzi2021}.  As such, asymmetry itself is not a guarantee of anisotropy.  Furthermore, the metamaterial properties achievable with locally-distorted asymmetric patterns are not likely to reflect on the entire possible anisotropic metamaterial property space \citep{Xu2016,Liu2019,Grima2008}.  A more general form of asymmetry, such as the asymmetric arrangement of lattice members, would allow for lattice patterns that are more likely to represent a broader range of anisotropy as well as unique combinations of anisotropic properties. Existing literature has yet to explore the benefits of an asymmetric design space of this nature.

One way to understand the lattice-based metamaterial design space is through bounds on the elastic moduli. There is a vast literature establishing bounds for mechanical properties of multiphase materials, some of which can readily be applied to lattice-based metamaterials for which one of the ``phases" of material is air. The variational approach laid out by Hashin and Shtrikman \citep{Hashin1963} was one of the earliest studies that defined bounds on elastic moduli without assuming any detailed knowledge about the microstructure of the material.  This study was predicated on relatively small ratios between the moduli of the two phases, and thus are not appropriate for lattice-based metamaterials. However, Hashin-Shtrikman bounds have since been extended and specified to many different cases. Multiphase composites were initially the focus of such work, much of which was conducted in the context of homogenization.  Early work in this field proved the existence of optimal bounds on bulk and shear moduli for composites with two solid phases \citep{Francfort1986,Milton1988} as well as that the elastic properties of such composites are bounded by the elastic properties of finite-rank laminates (ordered composites) \citep{Avellaneda1987}. Subsequently it was shown that any positive semi-definite elasticity tensor for a 2D composite can be realized by manipulating the microstructure of the two constituent materials of unequal stiffness \citep{Allaire1994,Silva1995,Burns1997,Graeme1995}. Homogenization has also been used as a tool for optimizing the topology of a wide variety of structures, but most notably truss-based structures \citep{Sigmund1994,Bendsoe1994}.  Building on the work regarding multiphase composites, Hashin-Shtrikman bounds have also been specialized for use with microstructures that include a void phase (such as lattice metamaterials). One of the key studies in this area developed general Hashin-Shtrikman bounds for any multiphase material with a void phase, in the context of homogenization \citep{Torquato1998}. One further specialization of this work was for open-cell and closed-cell foams modeled as closely-packed hollow spheres \citep{Grenestedt1999}. These upper bounds are well suited for use within linear elasticity. The lower bound of moduli in such materials is zero, which occurs when the microstructure of the solid phase is not fully interconnected.  This approach has been shown as suitable for ordered metamaterials at higher volume fractions \citep{DoRosario2017}.  Using a similar approach, Hashin-Shtrikman bounds for cellular solids consisting of multiple slender beams have also been developed \citep{Grenestedt1998}. These bounds are independent of the connectivity of beams, and therefore the degree of anisotropy present. As they model the microstructure as a network of beams, these bounds are appropriate for lattice metamaterials. The utility of these bounds were demonstrated by Gurtner and Durand \citep{Gurtner2014}, who found upper bounds on the moduli of the stiffest possible isotropic network of beams for any given density. 

A second way to understand the lattice-based metamaterial design space is by direct modeling. Modeling of lattice-based metamaterials is conducted through a variety of methods.  Metamaterials are typically modeled on the scale of a single unit cell, treating the single unit cell as if it resides within a broader continuum composed of the same unit cells.  The least computationally intensive method of modeling lattice patterns is through an analytical model.  These models represent the relationship between geometric parameters (such as member thickness or angles between members) and desired metamaterial properties for a specific lattice pattern \citep{Yu2018,Nicolaou2012,NaghaviZadeh2021}. Analytical models are useful in determining the degree to which individual input parameters influence desired metamaterial properties \citep{Cabras2016,Karathanasopoulos2017,Morita2021}.  However, the accuracy of a particular analytical model will depend on details of the lattice, and a single model may not be appropriate for a wide range of lattice patterns. On the other end of the spectrum, high-fidelity finite element analysis (FEA) methods are the most accurate and most computationally intensive method for determining the mechanical properties of lattice patterns.  Due to their use of 3D elements, these models are especially useful when studying lattice patterns with members of low length-to-thickness aspect ratios wherein the interaction of members at junctions has a significant effect on overall deformation characteristics \citep{Tancogne-Dejean2018,Tancogne-Dejean2019}.  However, high-fidelity FEA models are unnecessary when modeling metamaterials with sufficiently high length-to-thickness aspect ratios.  Reduced-order FEA models are more computationally efficient than high-fidelity models, and can be easily parameterized to work within a broad design space.  In such models, beam elements can accurately represent the behavior of the overall unit cell and the kinematics of junctions while sacrificing detail regarding the interaction of members at these junctions \citep{Asadpoure2015,Abdelhamid2018,Adhikari2021,Vangelatos2020,Jamshidian2020}.  Such beam elements are assumed to sustain normal and transverse loads, moments, and torques at junctions \citep{Bluhm2020,Mizzi2020,Liu2019}.  In situations where the enhanced accuracy due to the additional degrees of freedom present in beam elements is not as critical as reducing computational expense, truss elements can instead be used in reduced-order models \citep{Javadi2012,Asadpoure2015}.

In this study, the stiffness tensors of symmetric and asymmetric 2D lattice patterns with the same unit cell size are compared to determine the extent to which asymmetric lattice patterns allow for combinations of mechanical properties that are distinct from those offered by symmetric lattice structures.  First, random symmetric and asymmetric lattice patterns are generated.  The effective elastic stiffness tensor and volume fraction of all the lattice patterns are then calculated using a reduced-order finite element model.  The symmetric and asymmetric design spaces are presented and analyzed.  Designs in the asymmetric design space are shown to have distinct stiffness tensors from those in the symmetric design space.  The presence of certain design features is identified in many of the evaluated designs; these features are shown to have a statistically significant impact on the metamaterial properties of designs that contain them.  

\section{Methods}
\label{sec:methods}

\begin{figure}[hbt!]
    \centering
    \includegraphics[scale=0.4]{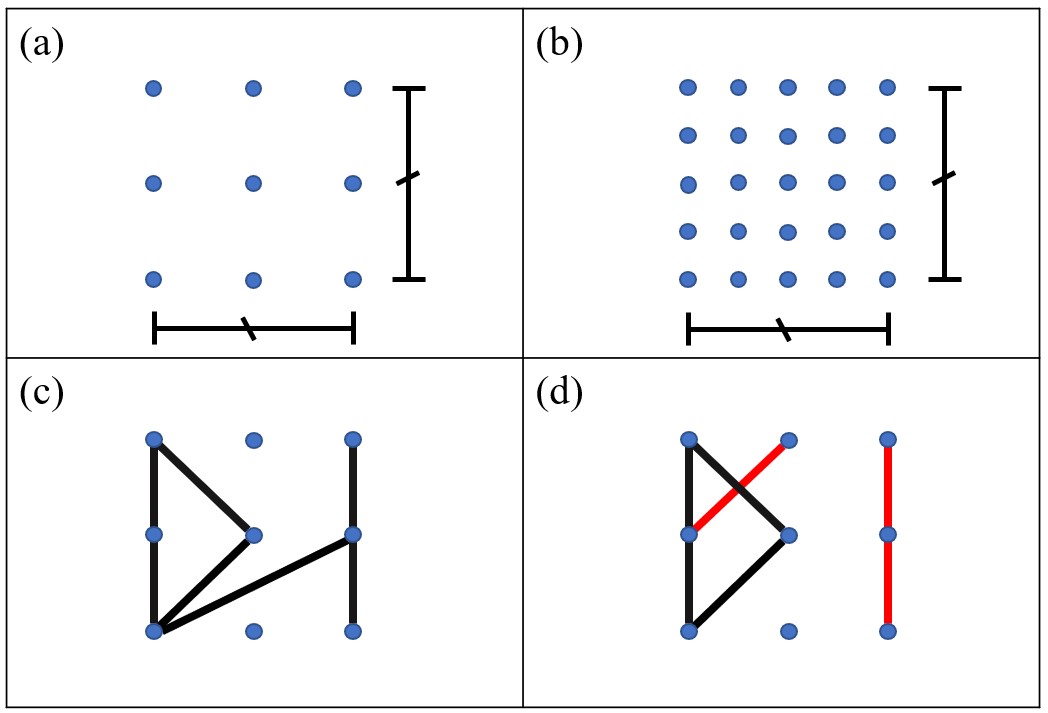}
    \caption{Definitions of the design spaces in this study.  The distribution of junctions are shown within the (a) 3x3 and (b) 5x5 design spaces.  As indicated in the figure, both of these design spaces are of equal outer dimension.  Shown in (c) is an example of a unit cell design that satisfies the feasibility criteria.  Shown in (d) is an example of a unit cell design that does not satisfy the feasibility criteria. The violating members are shown in red.  The first of these members intersects another member at a point outside of the junction grid, while the other two members at the right of the unit cell do not connect to the other half of the unit cell.}
    \label{fig:designspace}
\end{figure}

A standardized design space is used to study the difference between symmetric and asymmetric lattice patterns, such that the sole difference between symmetric and asymmetric patterns is the arrangement of members.  To facilitate visual identification of design traits, the design spaces used herein are restricted to 2D lattice patterns.  The underlying platform for all unit cell designs considered in this study is a grid of evenly-spaced junctions arranged in a square.  Two such design spaces are considered: a 3x3 grid and a 5x5 grid, as shown in Figures \ref{fig:designspace}a and b respectively.  Lattice members can span any two points on the grid, and all members are straight and have circular cross-sections with the same fixed radii.  The design space is constrained by four requirements to ensure that lattice patterns will form a continuous material, without any dangling members, that only has member intersections at junctions. Figures \ref{fig:designspace}c and d shows one design that passes and design that fails these constraints. These constraints are as follows:
\begin{enumerate}
    \item Designs cannot have any intersecting or overlapping members. 
    \item Designs cannot have any isolated members or groups of members.
    \item Designs must be connected to neighboring unit cells, to ensure that the design can tessellate. 
    \item All used junctions in a design should be connected to at least two other junctions in the grid.
\end{enumerate}

\begin{figure}[hbt!]
    \centering
    \includegraphics[scale=0.5]{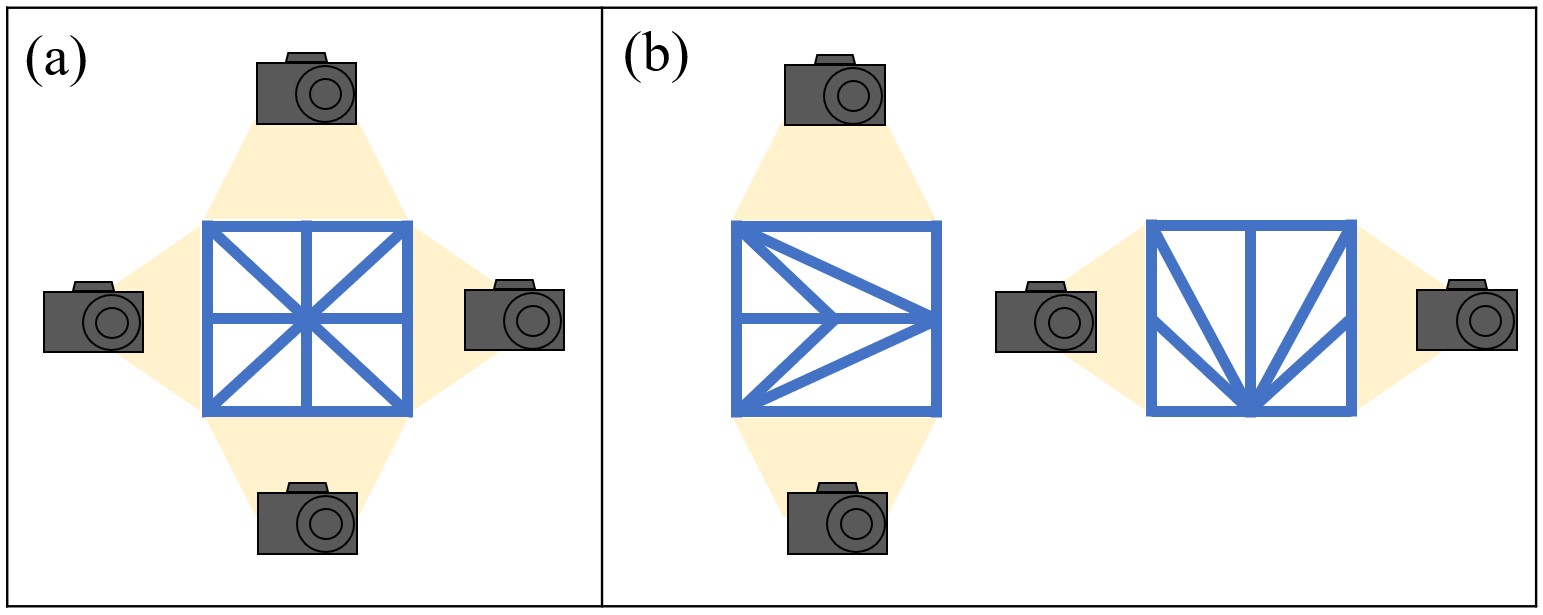}
    \caption{Definitions of symmetry terms in this study.  Double-mirror symmetry in the context of this study is shown in (a), wherein a design observed from all four sides appears the same.  Mirror symmetry is shown in (b), wherein a design observed from opposing sides, either vertically or horizontally, appears the same.}
    \label{fig:symmetrydef}
\end{figure}

Two types of symmetry are considered in this study: double-mirror and mirror (Figure \ref{fig:symmetrydef}b).  Unit cells with double-mirror symmetry appear identical when observed from all four sides. Unit cells with mirror symmetry appear identical when observed from opposite sides --- either vertically or horizontally --- but the horizontal and vertical views are different from each other.  Mirror symmetry does not necessarily result in a cubic stiffness tensor.  For this study, a design that satisfies neither of these definitions of symmetry is considered to be asymmetric (Figure \ref{fig:symmetrydef}a).

A generative design approach was used to develop sets of symmetric and asymmetric designs. Generative design is an emerging approach for generating design options, especially in the Computer Aided Design (CAD) community \citep{Krish2011}. The combination of generative design with additive manufacturing has been of great interest recently \citep{Wu2019} with the incorporation of surrogate modeling methods into the generative design framework explored in recent literature \citep{Oh2019,Yuan2020,Kallioras2020}. Valid designs in the dataset must satisfy the previously-discussed constraints on the design space.  Random assignment of members to the 2D grid space to generate the design dataset would result in the majority of designs not satisfying the constraints. In contrast, a generative design approach guarantees overall design feasibility by ensuring constraint satisfaction for every added member. Here, designs are generated by randomly adding feasible (non-intersecting and non-overlapping) members.  Additional reflected members are added as appropriate to incorporate symmetry (double-mirror or mirror). Each subsequent feasible member to be added is chosen randomly from a set of feasible members connected to the previously added member, with tessellation of the unit cell in all directions considered in identifying these members. In case no feasible members can be found connected to the previously added member, a feasible member is chosen from a set of feasible members connected to the node with the least number of connections. As a baseline, all members are assigned equal probability of selection. However, there is an additional provision to bias the probability of selecting a short member (members connected by nodes which are closest to each other in the vertical, horizontal or diagonal directions). This probability ranges from 0 (truly random selection of members) to 1 (only short members are added).  Varying this probability can alter the regions of the overall design space that generated designs occupy.  This generative approach ensures that satisfactorily feasible designs will be generated, while allowing us to target specific regions of the design space by biasing the probability of adding short members. After a satisfactory design is achieved, additional designs are created by adding feasible members to the satisfactory design until no more feasible members can be added. Then, a random subset of those designs is chosen to incorporate into the final dataset in order to overcome the bias towards sparse designs. Additionally, each generated design is also rotated in-plane by 90 degrees and included in the design set, provided it is not a repetition of an existing design.

The effective metamaterial elastic properties are determined for each design.  These elastic properties are ``effective" values because they represent the stiffness of an architected material rather than a homogeneous material.  The 2D stiffness tensor, found by modeling a single unit cell within an array of self-repeats, is shown in Equation 1:  
\begin{equation}
    \boldsymbol{\sigma} = \mathcal{C}:\boldsymbol{\varepsilon} = 
    \begin{bmatrix}
    \sigma_{11}\\
    \sigma_{22}\\
    \sigma_{12}
    \end{bmatrix} = 
    \begin{bmatrix}
    C_{11} & C_{12} & C_{16}\\
    C_{21} & C_{22} & C_{26}\\
    C_{61} & C_{62} & C_{66}
    \end{bmatrix}
    \begin{bmatrix}
    \varepsilon_{11}\\
    \varepsilon_{22}\\
    \varepsilon_{12}
    \end{bmatrix} ,
\end{equation}
\noindent where $C_{11}$ and $C_{22}$ are the two normal stiffness values and $C_{66}$ is the shear stiffness value.

The compliance tensor is calculated from the stiffness tensor as shown in Equation 2: 
\begin{equation}
    \mathcal{S} = \mathcal{C}^{-1} = 
    \begin{bmatrix}
    S_{11} & S_{12} & S_{16}\\
    S_{21} & S_{22} & S_{26}\\
    S_{61} & S_{62} & S_{66}
    \end{bmatrix} .
\end{equation}
The Poisson's ratios $\nu_{12}$ and $\nu_{21}$ are then calculated by Equations 3 and 4, assuming the orthotropy of the unit cells under small deformation.   
\begin{equation}
    \nu_{12} = -S_{12} \cdot C_{22} ,
\end{equation}
\begin{equation}
    \nu_{21} = -S_{21} \cdot C_{11} .
\end{equation}
In addition to the stiffness tensor, the volume fraction of each unit cell is calculated by finding the summed volume of all lattice members and dividing by the volume of a square panel taking up the same overall space as a single unit cell.  To explicitly account for the overlap caused by two members intersecting at a junction, the volume of a single overlap is approximated as the largest spherical wedge that can fit inside the space occupied by two members.  The angle of this wedge is defined as the span angle between the two intersecting members.  These overlap volumes are calculated for each intersection in a design, then cumulatively subtracted from the total volume of all lattice members in a design.

Bounds on elastic moduli provide context as to the theoretical limits of a design space.  For our work, we use previously-derived bounds for cellular solids that consist of elastic beams that meet at negligibly-sized joints \citep{Gurtner2014}.  The upper bounds on Young's modulus and shear modulus used herein are:
\begin{equation}
    \label{youngsbound}
    \frac{E^{HS}}{E_{0}} = \frac{\phi}{6}(1+3\frac{49\kappa_{0}+8\mu_{0}}{69\kappa_{0}+8\mu_{0}}) ,
\end{equation}
\begin{equation}
    \label{shearbound}
    \frac{\mu^{HS}}{E_{0}} = \frac{\phi}{9}(1+\frac{18\kappa_{0}^{2}+8\mu_{0}^{2}+21\kappa_{0}\mu_{0}}{3\kappa_{0}(3\kappa_{0}+4\mu_{0})}) ,
\end{equation}
\noindent wherein $\phi$ is volume fraction and $\kappa_{0}$ and $\mu_{0}$ are the bulk and shear moduli of the constituent material, respectively. For cellular solids, it has been shown that the lower bound on elastic moduli goes to zero \citep{Torquato1998}.  We use these equations to show the upper bounds on shear and Young's moduli as a function of $\phi$.

We take a reduced-order modeling approach to simulate a single unit cell of each lattice design as it would act within an array of self-repeats.  Two-node beam elements are used to represent each member; the beam type ascribed to such elements by ANSYS is a Timoshenko beam.  This model is written in the ANSYS Parametric Design Language (APDL) \citep{Canonsburg2010} and run in ANSYS Mechanical\textsuperscript{\texttrademark}.  All members have a radius $\frac{1}{40^{th}}$ the outer dimension of each square unit cell.  The number of junctions in the grid and the connectivity array of the design --- describing the position of each lattice member by which two junctions they connect --- are also inputs.  Periodic boundary conditions (PBCs) are used to simulate the behavior of a single unit cell within a larger continuum.  The PBCs are implemented through the use of dummy nodes, which are linked to points on the boundary of each unit cell. The outputs of each simulation are the reaction forces at these dummy nodes, which are used to determine the effective stress on the unit cell.  Three simulations with different applied strains are run to achieve the 9 components of the stiffness tensor for each lattice pattern.  The output result for each design is the metamaterial stiffness tensor, which is normalized relative to a constituent material Young's modulus.  MATLAB\textsuperscript{\texttrademark} is used to pre- and post-process the inputs and outputs for each design, and interfaces directly with ANSYS \citep{Zhan2018}.  All of our scripts are available on Github \citep{Srivatsa2021}.

Probability density plots are used to evaluate the distributions of individual components of stiffness tensors, in order to compare the metamaterial property spaces of designs with different types of symmetry.  These probability distributions cannot be assumed as normal if they fail a one-sample Kolmogorov-Smirnov (K-S) test, with a null hypothesis that each distribution can be represented by a normal distribution \citep{Pratt1981}. Non-normal distributions cannot be fully characterized using mean and variance.  Therefore, such distributions are plotted as kernel distributions with bandwidths calculated using the Sheather and Jones method to account for any multimodality present \citep{Hall1991}.  To recognize any differences between these kernel distributions, two-sample K-S tests are used with the null hypothesis that two probability distributions are from the same continuous distribution \citep{Pratt1981,Ibrahim2009}.  Two distributions that, when compared, reject the null hypothesis of a two-sample K-S test can be recognized as distinct from each other but cannot otherwise be quantified relative to each other.  Differences between two distinct distributions are thus discussed qualitatively.

\section{Results and Discussion}
\label{sec:resultsdisc}

In both the 3x3 and 5x5 design space, the symmetry-defined subsets of designs need to be both large enough to draw quantifiable conclusions regarding the distributions of metamaterial properties as well as broad enough to be representative of the design space.  Within each design space, we varied the probability of choosing a short member from 0 to 1 by increments of 0.2.  For each probability value, we started with an initial set of 50 designs and generated 50 designs at a time.  Before adding each additional batch to the overall set, we determine whether the newly-generated batch significantly changes the relative distribution of the existing set using a two-sample K-S test. We continue to add new batches of designs until the K-S test no longer rejects the null hypothesis (that the design set with the newest batch is distinct from the set without the newest batch).  Variations of this approach have been previously used in literature to understand the evolution of a data set based on its spread and relative density \citep{Seringhaus2006,Selk2013,MadridPadilla2019}. In the 3x3 design space, only 22 unique double-mirror symmetric designs were found to exist, so this population is sampled completely. Additionally, 300 unique 3x3 mirror symmetric designs and 400 unique asymmetric 3x3 designs were generated.  The 5x5 design space is large enough to facilitate more double-mirror symmetric designs. A 5x5 design set was generated consisting of 500 double-mirror symmetric designs, 750 mirror symmetric designs, and 800 asymmetric designs.  

\subsection{Metamaterial Properties as a Function of Symmetry}

\begin{figure}[hb!]
    \begin{center}
    \includegraphics[width=\textwidth]{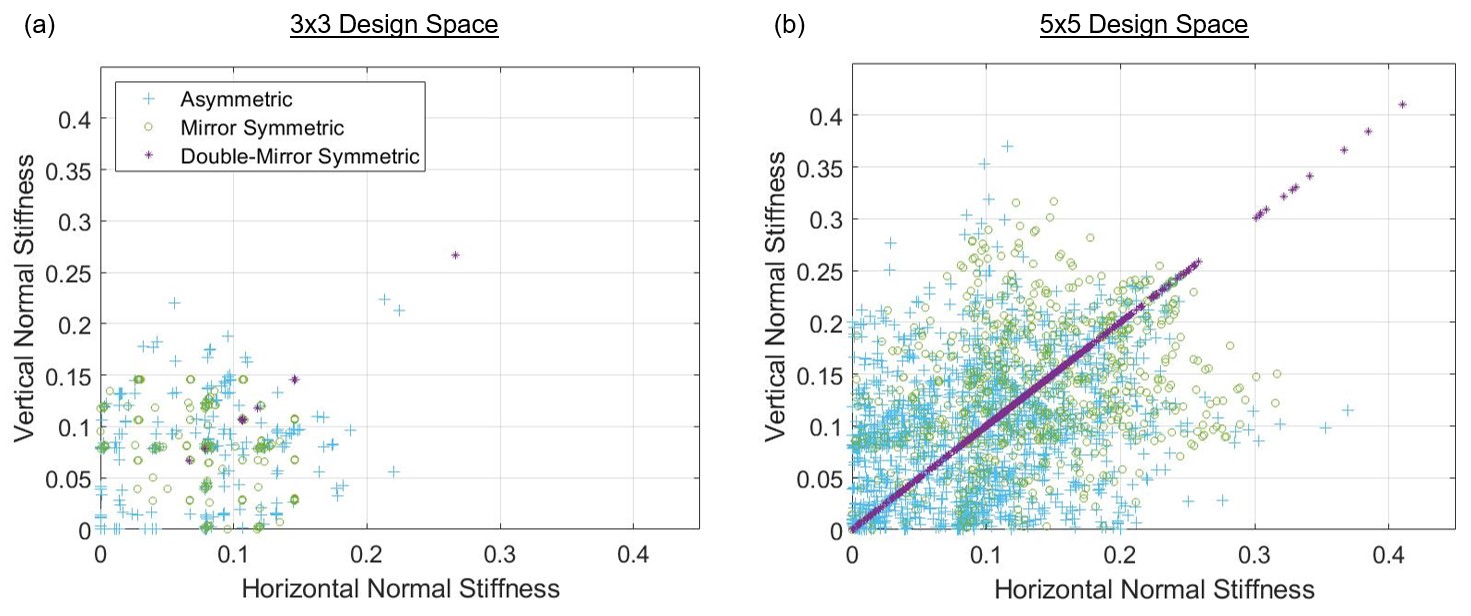}
    \end{center}
    \caption{Vertical and horizontal normal stiffness values compared against each other for (a) the 3x3 design space and (b) the 5x5 design space.}
    \label{fig:normalnormal}
\end{figure}

After finding the stiffness tensors for all designs, the normal stiffness values are examined.  The vertical and horizontal normal stiffness values for each design are plotted against each other in Figure \ref{fig:normalnormal}.  Double-mirror symmetric designs have the same horizontal and vertical normal stiffness as expected. Additionally, some asymmetric and mirror-symmetric designs also have normal stiffness values that are similar to each other.  Also evident from the design spaces is that the highest normal stiffness values belong to double-mirror symmetric designs. This is especially true in the 5x5 design space, in which double-mirror symmetric designs attain normal stiffness values exceeding 0.4 but only a few asymmetric and mirror symmetric designs have normal stiffness values over 0.3.

\begin{figure}[ht!]
    \begin{center}
    \includegraphics[width=\textwidth]{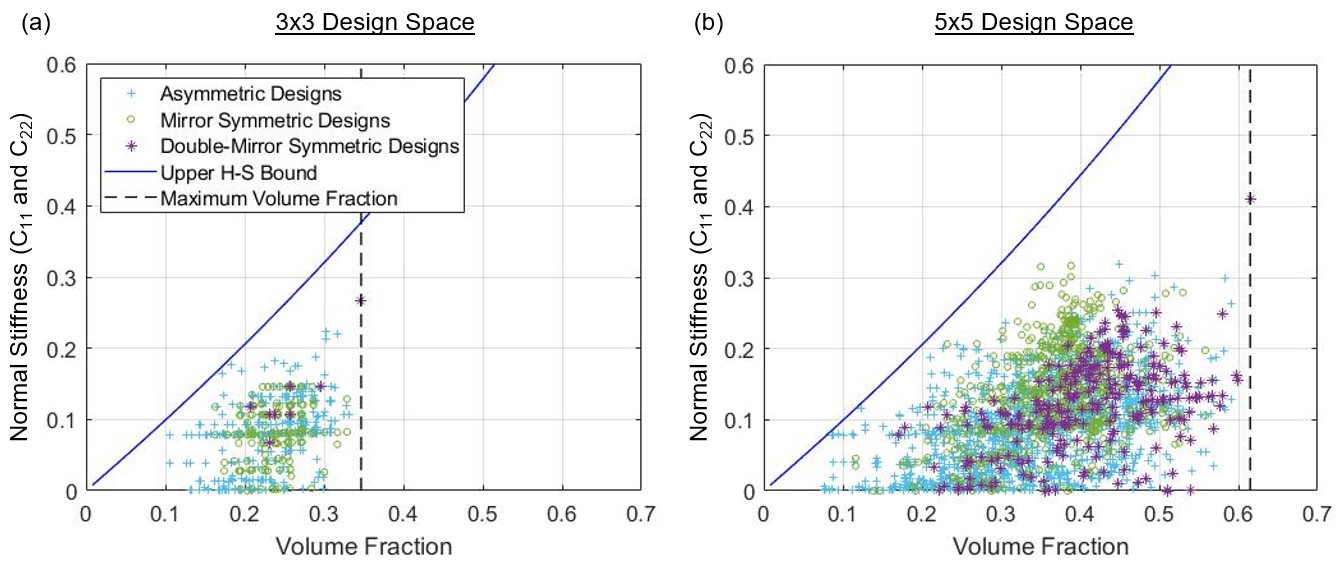}
    \end{center}
    \caption{Normal stiffness (both horizontal and vertical) compared against volume fraction for (a) the 3x3 design space and (b) the 5x5 design space. The solid lines in both plots are the Hashin-Shtrikman upper bounds (Eq. \ref{youngsbound}), and the dashed lines are the maximum achievable volume fraction in each design space. The Hashin-Shtrikman lower bound is 0.}
    \label{fig:normalvf}
\end{figure}

The normal stiffness values of designs are plotted against their volume fractions in Figure \ref{fig:normalvf}.  As expected, the higher resolution of 5x5 designs results in these designs achieving higher volume fractions and higher stiffness than 3x3 designs.  Additionally, all designs in both design sets are bounded within Hashin-Shtrikman bounds specialized for lattice-type solids \citep{Grenestedt1999}.  However, even the most dense design in the 5x5 design space has a normal stiffness much less than the upper-bound stiffness value at that density.  The stiffer designs that may exist at that density would likely be outside the constraints of the prescribed design space (a nuance not captured by the bounds used).  There is a positive correlation between volume fraction and normal stiffness in both design spaces: for all six subsets the slopes of linear fits range from 0.310 to 0.578.  This is expected, as unit cells with more members are generally expected to be stiffer.  However, designs with similar volume fractions can vary significantly in normal stiffness, beyond even an order of magnitude; R-squared values of the linear fits range from 0.139 to 0.378. 

\begin{figure}[ht!]
    \begin{center}
    \includegraphics[width=\textwidth]{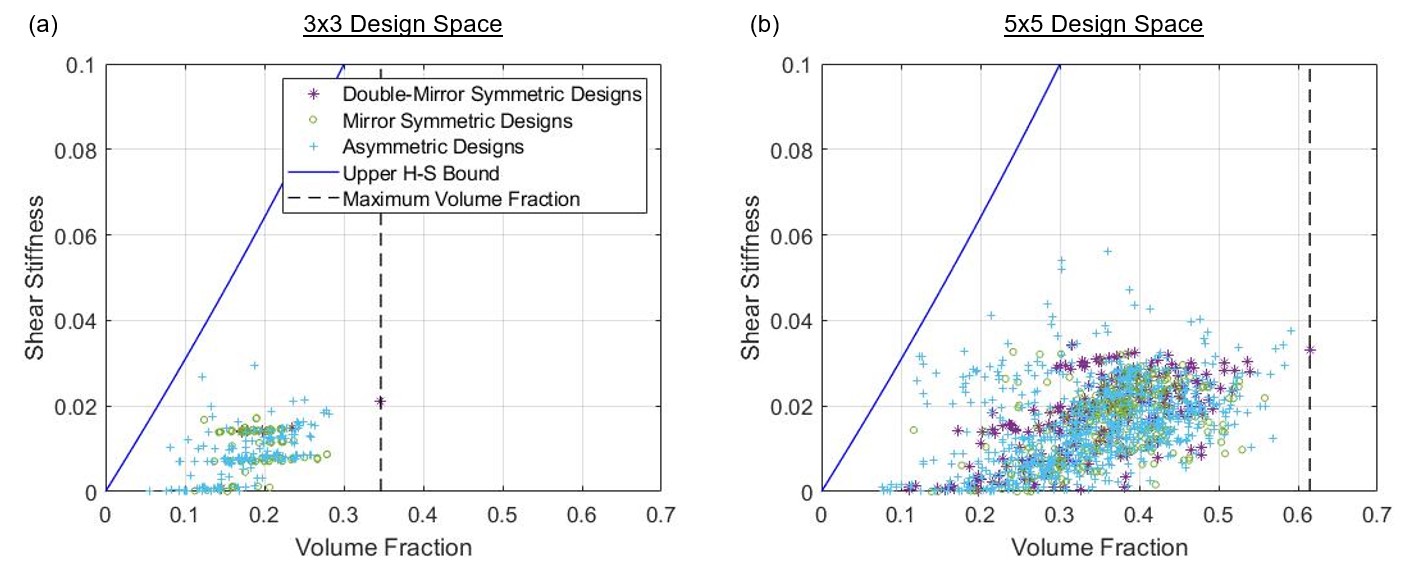}
    \end{center}
    \caption{Shear stiffness compared against volume fraction (a) the 3x3 design space and (b) the 5x5 design space. The solid lines in both plots are the Hashin-Shtrikman upper bounds (Eq. \ref{shearbound}), and the dashed lines are the maximum achievable volume fraction in each design space. The Hashin-Shtrikman lower bound is 0.}
    \label{fig:shearvf}
\end{figure}

The shear stiffness values of designs are plotted against their volume fractions in Figure \ref{fig:shearvf}.  All designs in both design spaces lie again within Hashin-Shtrikman bounds.  Unlike with normal stiffness, there is only a slightly positive correlation between volume fraction and shear stiffness in both design spaces.  The slopes of linear fits range from 0.0387 to 0.0950, with R-squared values ranging from 0.0845 to 0.428. As with the normal stiffness values, the shear stiffness values are further below the corresponding upper bound at higher volume fractions than at lower volume fractions. Asymmetric designs have the largest shear stiffness of the three symmetry sets; this will be explored further in Section \ref{subsect:DesFeatures}.  Wilcoxon rank sum tests \citep{Wilcoxon1945} found that the shear stiffness values of asymmetric designs were distinct from those of mirror symmetric designs (U = 433638 and p < 0.001 in 3x3, U = 247550 and p < 0.001 in 5x5) and double-mirror symmetric designs (U = 6291156 and p < 0.001 in 3x3, U = 7008216 and p < 0.001 in 5x5).  Because these values are less than a significance level of 0.05/3 (adjusted using a Bonferroni correction to mitigate the sampling of the same distribution for multiple tests \citep{Bonferroni1936}), the population of asymmetric designs has a distinct median shear stiffness from the populations of mirror and double-mirror symmetric designs. There are some asymmetric designs at lower volume fractions that have higher normalized shear stiffness values than any double-mirror symmetric designs. The higher shear stiffness-to-volume fraction ratios in asymmetric designs is potentially due to more diagonally aligned members allowed by asymmetry, which is an example of how symmetry can constrain the material property space.

\begin{figure}[ht!]
    \begin{center}
    \includegraphics[width=\textwidth]{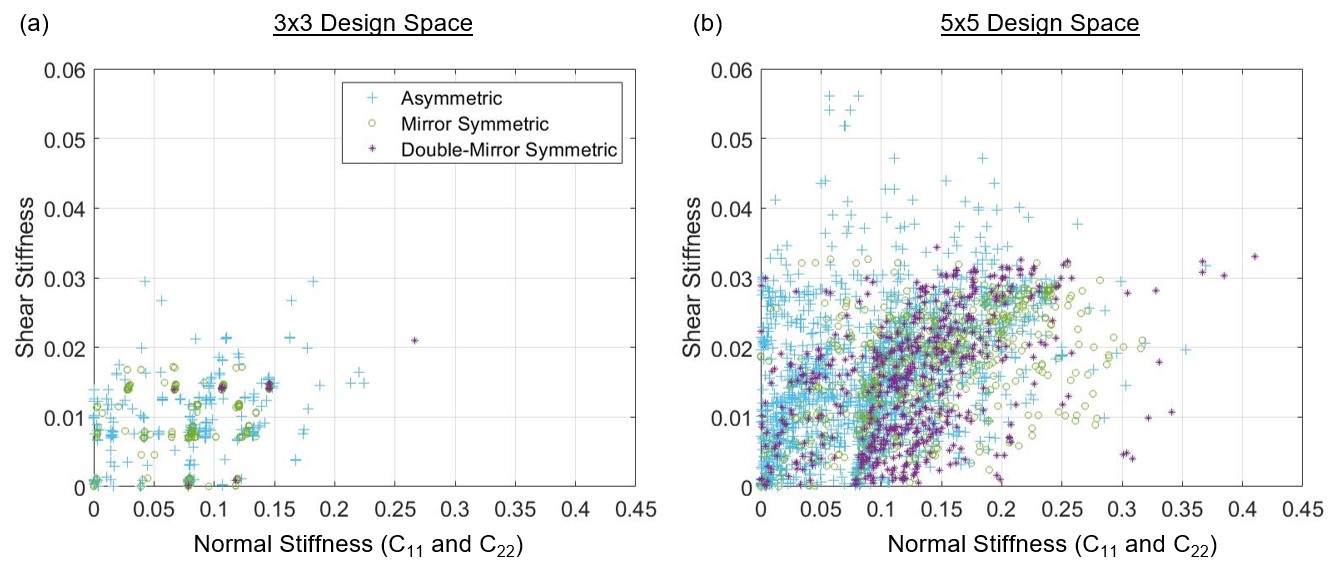}
    \end{center}
    \caption{Shear stiffness compared against normal stiffness for (a) the 3x3 design space and (b) the 5x5 design space.}
    \label{fig:normalshear}
\end{figure}

Shown in Figure \ref{fig:normalshear} is the shear stiffness of each design plotted against its normal stiffness.  Designs of any symmetry type are only slightly more likely to have a larger shear stiffness if they have a larger normal stiffness.  This weak positive correlation is supported by the relatively low slopes (varying from 0.0317 to 0.103) and R-squared values (varying from 0.0756 to 0.485). Interestingly, many of the shear stiffness values are less than the $\frac{1}{3^{rd}}$ minimum ratio of shear to elastic modulus required for an isotropic solid. Clustering of metamaterial properties is more prominent in the 3x3 design space; the additional resolution of the 5x5 design space allows for a broader array of lattice patterns and thus a more diverse metamaterial property space. 

\begin{figure}[ht!]
    \begin{center}
    \includegraphics[width=\textwidth]{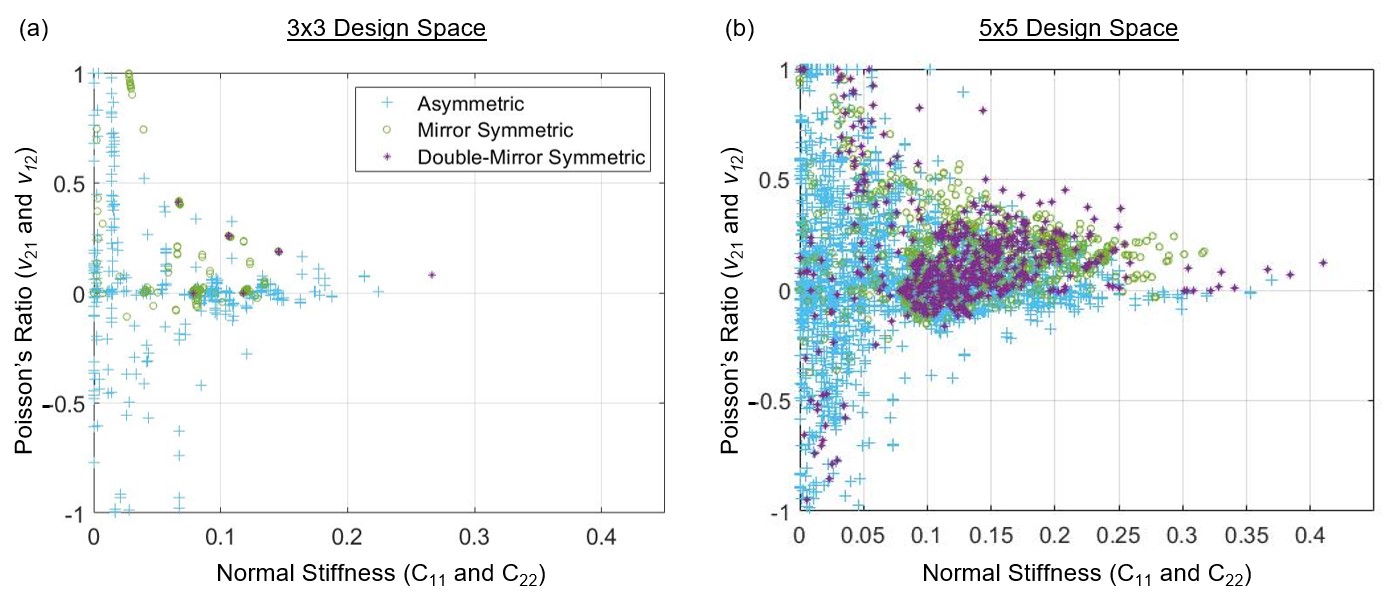}
    \end{center}
    \caption{Poisson's ratio compared against normal stiffness for (a) the 3x3 design space and (b) the 5x5 design space.} 
    \label{fig:normalpoisson}
\end{figure}

Next the Poisson's ratios ($\nu_{12}$, $\nu_{21}$) are examined. Figure \ref{fig:normalpoisson} shows the Poisson's ratio associated with deformation transverse to the plotted normal stiffness direction. Within the design space sampled, the boundaries of the metamaterial property space exhibit an inverse relationship between the normal stiffness of a design and the magnitude of the associated Poisson's ratio. Within the 3x3 design space, greater asymmetry corresponds to lower average Poisson's ratio and larger maximum-minimum range in values (means of 0.0231, 0.136, 0.207 and ranges of 1.99, 1.11, 0.414 for asymmetric, mirror symmetric, and double-mirror symmetric designs respectively). In the less constrained 5x5 design space, the ranges of the three design sets are similar to each other. However, the mean of the fully asymmetric design set remains lower than the other two (means of 0.0623, 0.180, 0.146 and ranges of 1.99, 1.95, 1.97 for asymmetric, mirror symmetric, and double-mirror symmetric designs respectively). Asymmetry therefore increases access to negative Poisson's ratios. Nonetheless, the greater resolution of the 5x5 design space reduces the impact of symmetry on achievable Poisson's ratios.

\begin{figure}[ht!]
    \begin{center}
    \includegraphics[width=\textwidth]{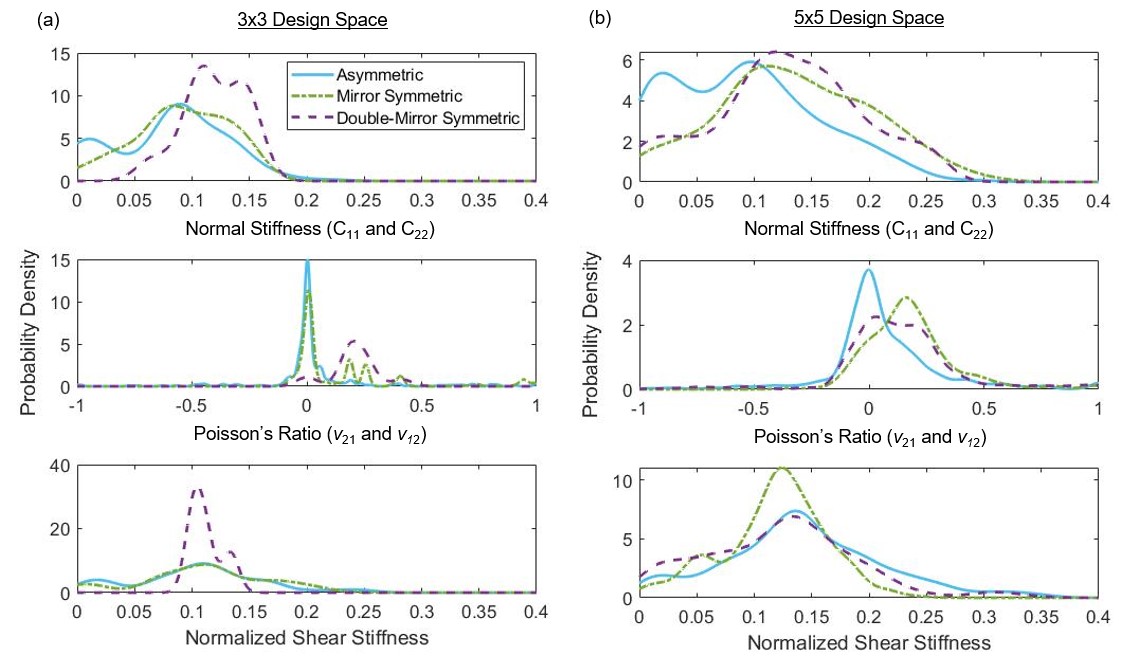}
    \end{center}
    \caption{Probability density plots comparing normal stiffness, Poisson's ratios, and normalized shear stiffness of asymmetric, mirror symmetric, and double-mirror symmetric designs in the (a) 3x3 and (b) 5x5 design spaces.  Shear stiffness values are normalized by the average normal stiffness of that design.}
    \label{fig:aggregatepdfs}
\end{figure}

In order to compare the metamaterial properties of asymmetric, mirror symmetric, and double-mirror symmetric designs, probability density distributions of these properties were generated (Figure \ref{fig:aggregatepdfs}).  All of these distributions were found to not be normal, and are thus represented using kernel distributions.  The results of the statistical tests used to determine non-normality, as well as the bandwidths of the kernel distributions in Figure \ref{fig:aggregatepdfs}, are included in the Supplementary Information.  Two-sample K-S tests were used to compare asymmetric, mirror symmetric, and double-mirror symmetric distributions to each other for distributions of normal stiffness, Poisson's ratio, and shear stiffness normalized by normal stiffness.  All of these tests return a p-value below p < 0.001, which is lower than a significance level of 0.05/3 (due to the Bonferroni correction required for multiple statistical tests performed on the same set \citep{Bonferroni1936}). These results show that the distributions of these three metamaterial properties for asymmetric, mirror symmetric, and double-mirror symmetric designs are all distinct from each other. These distributions also indicate how the mere presence of designs in a portion of the design space is not inherently indicative of relative density.  For instance, while Figure \ref{fig:normalnormal}b shows that only double-mirror symmetric designs in the 5x5 design space have the largest achievable normal stiffness values, the probability distribution shows that designs with the largest normal stiffness values are only a small subset of the design space.

\subsection{Design Features}

\begin{figure}[ht!]
    \begin{center}
    \includegraphics[width=\textwidth]{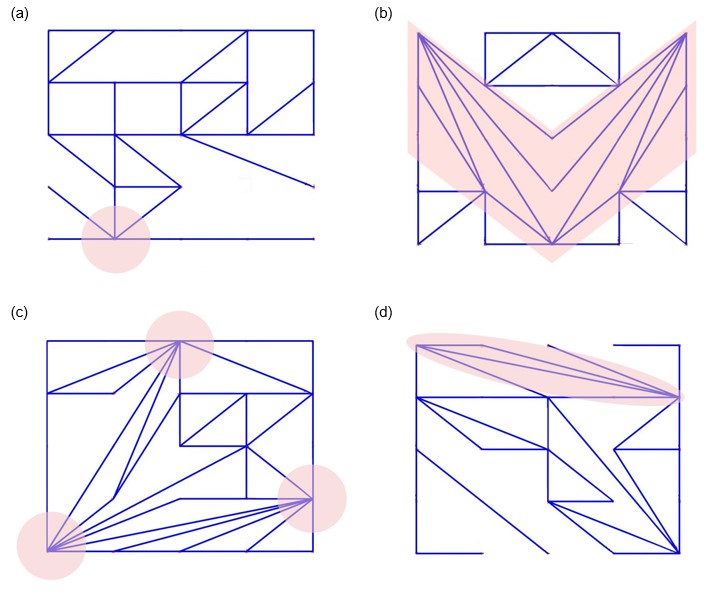}
    \end{center}
    \caption{Design features.  A \textit{pivot point} is shown in (a), wherein the lowest tier of members would be able to rotate relative to the rest of the design.  Two \textit{arrows} are highlighted in (b) with one stacked on top of the other.  Three \textit{spider nodes} are shown in (c), while a grouping of \textit{stacked members} is highlighted in (d).}
    \label{fig:samplefeatures}
\end{figure}

To better understand the differences in these design spaces and what mechanisms are responsible for the distinctions between the property probability distributions, four recurring design features were identified visually by qualitatively examining a random sample of the generated designs (Figure \ref{fig:samplefeatures}). These four features, chosen to represent relatively distinct patterns observed in designs, are \textit{pivot points}, \textit{arrows}, \textit{spider nodes}, and \textit{stacked members}.  A \textit{pivot point} is defined as a single node about which one portion of a design can rotate relative to the rest of the design, if the lattice pattern were to consist of pin joints and rigid members.  Even though the members have fully bonded intersections at junctions, the length-to-radius ratio of the members is large enough to promote member bending and thus the lattice patterns are flexible enough to experience nonaffine deformations if \textit{pivot points} are present.  \textit{Arrows} have been documented as a versatile lattice shape to achieve negative Poisson's ratios \citep{Qiao2015,Wang2018a}.  These shapes are defined as a grouping of four members that form a quadrilateral, with three acute angles and one angle over 180 degrees.  Similar to \textit{pivot points}, \textit{arrows} allow for relative movement within a lattice unit cell.  The third type of feature present, \textit{spider nodes}, are defined as junctions that have at least 6 connecting members as well as at least one pair of adjacent connecting members with no more than 45 degrees between them.  \textit{Stacked members} are defined as a group of at least three members that deviate in angle from each other no more than 10 degrees, terminating at a single point on either end of the group.  \textit{Spider nodes} and \textit{stacked members} both indicate a concentration of members that are at an angle, which could potentially contribute to increasing the shear stiffness of a given design. While the presence of any four of these features is inherently independent of the presence of the other three, it is possible for the same group of members to form multiple features.  We find the probability distributions of metamaterial properties for designs with and without these features to help assess to what extent, if any, these features impact the metamaterial property space.

\begin{figure}[ht!]
	\begin{center}
    \includegraphics[width=\textwidth]{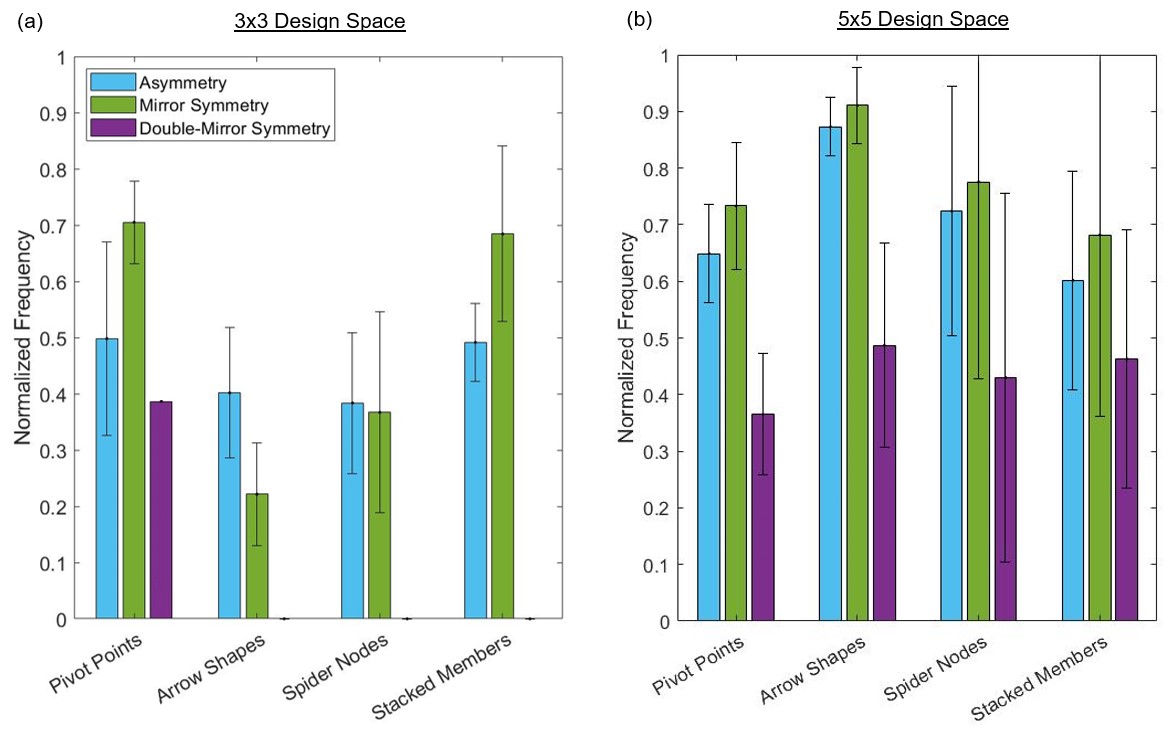}
	\end{center}
	\caption{Average normalized frequency of design features in asymmetric, mirror symmetric, and double-mirror symmetric designs with (a) 3x3 and (b) 5x5 nodal grids.  For each type of symmetry, the frequency of design features was found for 15 random sets, then averaged over all sets. Error bars indicate one standard deviation in normalized frequency for each set.  The exception is the 3x3 double-mirror symmetric design set; this design subspace was sampled completely so it was not divided into random groups.}
	\label{fig:featuresbars}
\end{figure}

Designs were first divided by their type of symmetry (asymmetric, mirror symmetric, and double-mirror symmetric), and then each of these symmetry-defined subsets was further divided into 15 random groups.  Each of the asymmetric and mirror symmetric groups of 3x3 designs consisted of 20 designs and 26-27 designs, respectively.  Similarly, the asymmetric, mirror symmetric, and double mirror symmetric groups of 5x5 designs each consisted of 33-34 designs, 50 designs, and 53-54 designs respectively.  Within each of these 45 groups, the fraction of designs with each design feature was calculated.  Designs were considered to have a feature if at least one occurrence of that feature was present.  Within each symmetry subset, the mean and standard deviation of all 15 proportions was then found (Figure \ref{fig:featuresbars}). The only design features seen in double-mirror symmetric designs in the 3x3 design space are \textit{pivot points}. These findings indicate that asymmetry is generally more conducive to the inclusion of these four features than double-mirror symmetry, but not as conducive as mirror symmetry. This is reasonable since the double-mirror symmetric space is highly restricted in such a way that reduces accessibility of most of these features, whereas a single mirror direction can promote features, for example arrows and pivots with the critical junctions on the central line. 



\subsection{Influence of Design Features on Metamaterial Mechanical Properties}
\label{subsect:DesFeatures}
Having identified the asymmetric, mirror symmetric, and double-mirror symmetric designs with each type of design feature, the correlation of these features with metamaterial mechanical properties can be assessed.  Normal stiffness, shear stiffness normalized by average normal stiffness, and Poisson's ratios were grouped based on whether or not a design had a given design feature.  This grouping resulted in two probability distributions for each metamaterial property.  Kernel distributions were used to represent each subset, as none of the subsets were found to be normally-distributed.  Each pair of distributions were compared using two-sample K-S tests, with the null hypothesis that these two distributions are from the same continuous distribution.  By conducting this process for each of the three mechanical properties, each of the four design features, all three types of symmetry, and in both the 3x3 and 5x5 design spaces, a total of 72 pairs of probability distributions were generated and compared to each other.

\begin{table}[ht!]
{{\scriptsize
\caption{Two-Sample K-S Test results for normal stiffness, Poisson's ratio, and shear stress normalized by normal stress.  Values shown are the p-values associated with each test.  A p-value less than 0.05, indicating rejection of the null hypothesis, is marked with an asterisk.  Arrows, spider nodes, and stacked members are not present in double-mirror symmetric designs in the 3x3 design space.  Cells without a K-S test result are for distributions where the design space does not contain any of the stated features.}
\label{tbl:comparefeatures}
\begin{center}
\begin{tabular}{ |c|c|c|c|c|c| }
 \hline
 \textbf{\begin{tabular}{@{}c@{}}Design \\ Space\end{tabular}} & \textbf{\begin{tabular}{@{}c@{}}Design \\ Feature\end{tabular}} & \textbf{\begin{tabular}{@{}c@{}}Mechanical \\ Property\end{tabular}} & \textbf{Asymmetric} & \textbf{\begin{tabular}{@{}c@{}}Mirror \\ Symmetric\end{tabular}} &  \textbf{\begin{tabular}{@{}c@{}}Double-Mirror \\ Symmetric\end{tabular}}\\ 
 \hline
 \textbf{3x3} & \textbf{Pivots} & \textbf{Normal} & $< 0.001^{*}$ & 0.0758 & $< 0.001^{*}$ \\ 
 \hline
 \textbf{3x3} & \textbf{Pivots} & \textbf{Poisson's} & $0.0264^{*}$ & $< 0.001^{*}$ & 0.0817 \\ 
 \hline
 \textbf{3x3} & \textbf{Pivots} & \textbf{Shear} & $< 0.001^{*}$ & $< 0.001^{*}$ & $< 0.001^{*}$ \\ 
 \hline
 \textbf{3x3} &  \textbf{Arrows} & \textbf{Normal} & $< 0.001^{*}$ & $< 0.001^{*}$ & - \\  
 \hline
 \textbf{3x3} &  \textbf{Arrows} & \textbf{Poisson's} & $< 0.001^{*}$ & $< 0.001^{*}$ & - \\  
 \hline
 \textbf{3x3} &  \textbf{Arrows} & \textbf{Shear} & $< 0.001^{*}$ & $< 0.001^{*}$ & - \\  
 \hline
 \textbf{3x3} &  \textbf{Spiders} & \textbf{Normal} & $< 0.001^{*}$ & $< 0.001^{*}$ & - \\  
 \hline
 \textbf{3x3} &  \textbf{Spiders} & \textbf{Poisson's} & $< 0.001^{*}$ & $< 0.001^{*}$ & - \\  
 \hline
 \textbf{3x3} &  \textbf{Spiders} & \textbf{Shear} & $< 0.001^{*}$ & $< 0.001^{*}$ & - \\  
 \hline
 \textbf{3x3} &  \textbf{Stacks} & \textbf{Normal} & 0.118 & $< 0.001^{*}$ & - \\
 \hline
 \textbf{3x3} &  \textbf{Stacks} & \textbf{Poisson's} & 0.0923 & $< 0.001^{*}$ & - \\  
 \hline
 \textbf{3x3} &  \textbf{Stacks} & \textbf{Shear} & $< 0.001^{*}$ & $< 0.001^{*}$ & - \\  
 \hline
 \textbf{5x5} &  \textbf{Pivots} & \textbf{Normal} & $0.0104^{*}$ & $0.0129^{*}$ & 0.216 \\ 
 \hline
 \textbf{5x5} &  \textbf{Pivots} & \textbf{Poisson's} & $< 0.001^{*}$ & 0.470 & 0.799 \\ 
 \hline
 \textbf{5x5} &  \textbf{Pivots} & \textbf{Shear} & $< 0.001^{*}$ & 0.515 & 0.0798 \\ 
 \hline
 \textbf{5x5} &  \textbf{Arrows} & \textbf{Normal} & $< 0.001^{*}$ & $< 0.001^{*}$ & $< 0.001^{*}$ \\ 
 \hline
 \textbf{5x5} &  \textbf{Arrows} & \textbf{Poisson's} & $< 0.001^{*}$ & $< 0.001^{*}$ & $< 0.001^{*}$ \\  
 \hline
 \textbf{5x5} &  \textbf{Arrows} & \textbf{Shear} & $< 0.001^{*}$ & $< 0.001^{*}$ & $< 0.001^{*}$ \\  
 \hline
 \textbf{5x5} &  \textbf{Spiders} & \textbf{Normal} & $< 0.001^{*}$ & $< 0.001^{*}$ & $< 0.001^{*}$ \\  
 \hline
 \textbf{5x5} &  \textbf{Spiders} & \textbf{Poisson's} & $< 0.001^{*}$ & 0.544 & $< 0.001^{*}$ \\  
 \hline
 \textbf{5x5} &  \textbf{Spiders} & \textbf{Shear} & $< 0.001^{*}$ & $< 0.001^{*}$ & $< 0.001^{*}$ \\  
 \hline
 \textbf{5x5} &  \textbf{Stacks} & \textbf{Normal} & $< 0.001^{*}$ & $< 0.001^{*}$ & 0.571 \\ 
 \hline
 \textbf{5x5} &  \textbf{Stacks} & \textbf{Poisson's} & $< 0.001^{*}$ & $< 0.001^{*}$ & $< 0.001^{*}$ \\  
 \hline
 \textbf{5x5} &  \textbf{Stacks} & \textbf{Shear} & $< 0.001^{*}$ & $0.0362^{*}$ & $< 0.001^{*}$ \\  
 \hline
\end{tabular}\\
\end{center}
}}\end{table}

The results of all 72 two-sample K-S tests are shown in Table \ref{tbl:comparefeatures}.  All four features appear to be consistently influential on the material property space; only two pairs of distributions in the asymmetric design space, four pairs in the mirror symmetric design space, and three pairs in the double-mirror symmetric design space are not statistically distinct.  However, there are more pairs of distinct distributions in the asymmetric design space and the mirror symmetric space than the double-mirror symmetric space.  In addition to the asymmetric and mirror symmetric design spaces being more conducive to the mere presence of features than the double-mirror design space, these results indicate that features present in these two design spaces are more likely to have an influence on metamaterial properties.  Below we focus our discussion on two sets of distributions - the 5x5 design space for \textit{arrows} as they influence Poisson's ratio and \textit{spider nodes} as they influence shear stiffness; the rest of the distributions are included in the Supplementary Information.

\begin{figure}[ht!]
    \centering
    \includegraphics[width=\textwidth]{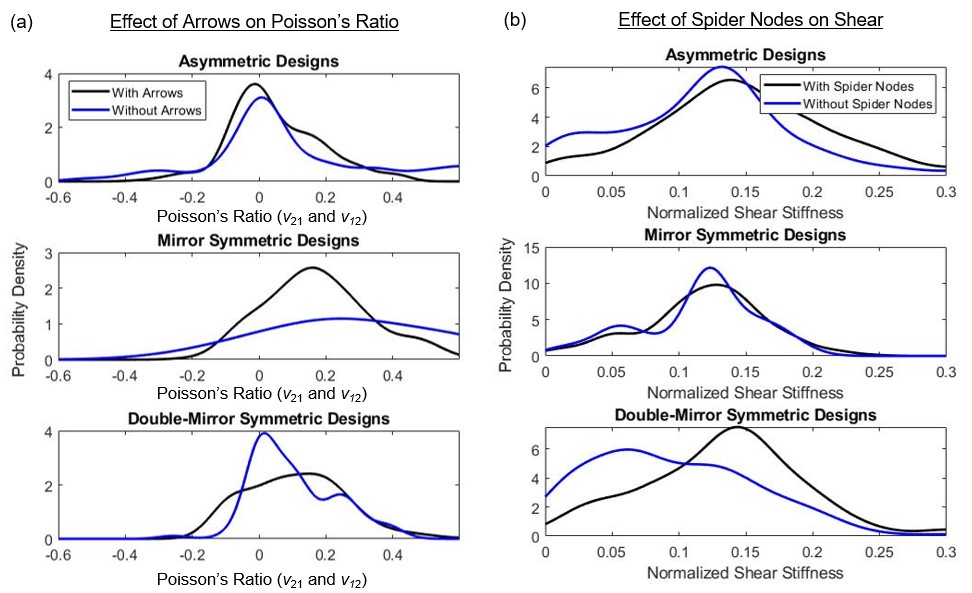}
    \caption{Probability density distributions of (a) Poisson's ratio for designs with and without \textit{arrows}, and (b) shear stiffness normalized by normal stiffness for designs with and without \textit{spider nodes}.  Both sets of distributions are in the 5x5 design space. All pairs of distributions are statistically distinct according to the two-sample K-S test, albeit to different degrees (refer to p-values in Table \ref{tbl:comparefeatures}).}
    \label{fig:comparefeatures}
\end{figure}

Shown in Figure \ref{fig:comparefeatures}a are the probability distributions of Poisson's ratio for designs with and without \textit{arrows}.  Figure \ref{fig:comparefeatures}b shows the probability distributions of normalized shear stiffness for designs with and without \textit{spider nodes}.  Both sets of distributions cover all three types of symmetry in the 5x5 design space.  All pairs of distributions shown are statistically distinct from each other. This is expected, as it was thought that \textit{arrows} and \textit{spider nodes} would influence the Poisson's ratios and normalized shear stiffness of designs, respectively.  However, the manner in which these features impact the property spaces varies. Because \textit{arrows} were thought to introduce a higher degree of relative movement within a unit cell, and thus more negative Poisson's ratios for one loading direction, the probability distribution of designs with \textit{arrows} is expected to be broader than the distribution of designs without \textit{arrows}.  Likewise, \textit{spider nodes} were thought to introduce a concentration of members and thus an increased normalized shear stiffness; the probability distributions with \textit{spider nodes} are expected to be shifted to the right of those without \textit{spider nodes}.  In both Figure \ref{fig:comparefeatures}a and \ref{fig:comparefeatures}b, these expected differences between distributions are seen only for asymmetric and double-mirror symmetric designs.  We verified these observations by finding the variances of the distributions in Figure \ref{fig:comparefeatures}a and the means of those in Figure \ref{fig:comparefeatures}b.  We then used two-sample F-tests to prove a significant difference of variance and two-sample t-tests to prove significantly distinct means (expecting p-values below 0.05 in both cases to affirm significance). The variances of the distributions with \textit{arrows} are significantly greater than for the distributions without \textit{arrows} for both asymmetric and double-mirror symmetric designs (Table \ref{tbl:arrowsvars}). Similarly, the means of the distributions with \textit{spider nodes} (Table \ref{tbl:spidersmeans}) are significantly greater than for the distributions without \textit{spider nodes} for both asymmetric and double-mirror symmetric designs (with p-values of 0.0289 and 0.0301 respectively).  Mirror-symmetric distributions in both cases do not conform to these expected outcomes, despite both features being most prevalent in mirror-symmetric designs.  These results indicate that the mere prevalence of a feature in a design space does not determine whether that feature has the intended impact on the property space.

\begin{table}
{{\scriptsize
\caption{Variances of probability distributions shown in Figure \ref{fig:comparefeatures}a.  Shown in the last column are the p-value results of the two-sample F-tests; statistically significant results are marked with an asterisk.}
\label{tbl:arrowsvars}
\begin{center}
\begin{tabular}{ |c|c|c|c| } 
 \hline
 \textbf{\begin{tabular}{@{}c@{}}Symmetry Type\end{tabular}} & \textbf{\begin{tabular}{@{}c@{}}Variance \\ with \\ Feature\end{tabular}} & \textbf{\begin{tabular}{@{}c@{}}Variance \\ without \\ Feature\end{tabular}} &
 \textbf{\begin{tabular}{@{}c@{}}p-value\end{tabular}}\\
 \hline
 \textbf{Asymmetric} &  0.0935 &  0.0712 & $< 0.001^{*}$\\
 \hline
 \textbf{Mirror Symmetric} &  0.0819 &  0.246 & $< 0.001^{*}$\\
 \hline
 \textbf{Double-Mirror Symmetric} &  0.128 &  0.0721 & $< 0.001^{*}$\\
 \hline
\end{tabular}\\
\end{center}
}}\end{table}

\begin{table}
{{\scriptsize
\caption{Means of probability distributions shown in Figure \ref{fig:comparefeatures}b.  Shown in the last column are the p-value results of the two-sample t-tests; statistically significant results are marked with an asterisk.}
\label{tbl:spidersmeans}
\begin{center}
\begin{tabular}{ |c|c|c|c| } 
 \hline
 \textbf{\begin{tabular}{@{}c@{}}Symmetry Type\end{tabular}} & \textbf{\begin{tabular}{@{}c@{}}Mean \\ with \\ Feature\end{tabular}} & \textbf{\begin{tabular}{@{}c@{}}Mean \\ without \\ Feature\end{tabular}} &
 \textbf{\begin{tabular}{@{}c@{}}p-value\end{tabular}}\\
 \hline
 \textbf{Asymmetric} &  0.184 &  0.158 & $< 0.001^{*}$\\
 \hline
 \textbf{Mirror Symmetric} &  0.132 &  0.137 & $< 0.001^{*}$\\  
 \hline
 \textbf{Double-Mirror Symmetric} &  0.148 &  0.128 & $< 0.001^{*}$\\
 \hline
\end{tabular}\\
\end{center}
}}\end{table}

\begin{figure}[ht!]
    \centering
    \includegraphics[width=\textwidth]{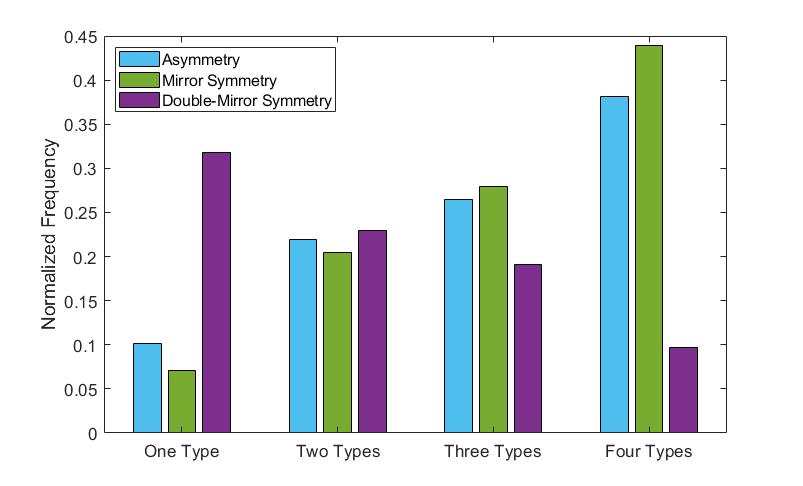}
    \caption{Normalized frequency of the number of types of features present in the 5x5 asymmetric, mirror symmetric, and double-mirror symmetric design sets used in this study.  The categories represent the number of different feature types a design has.  For each feature type, designs have at least one instance of that feature.}
    \label{fig:5x5_desChar_intersections_simple}
\end{figure}

A potential explanation for this discrepancy is found by examining the presence of multiple features in designs.  Figure \ref{fig:5x5_desChar_intersections_simple} shows the normalized frequency of designs of each symmetry category that have only one, any two, any three, or all four types of features.  Double-mirror symmetric designs are by far the most common category to have only one of the four features.  However, as the number of feature types present increases, asymmetric and mirror-symmetric designs become more common to the point that they are much more likely than double-mirror symmetric designs to have all four feature types present. Asymmetric designs, which do not have any symmetry constraints, likely can accommodate multiple features without an overlap that causes features to influence each others' performance. The mirror symmetric design space has similar feature-type proportions to the asymmetric design space, but also a lack of the intended impact of these features (Figure \ref{fig:comparefeatures}). The requirements of mirror symmetry limit the space available for features to exist in designs; different features are likely to overlap because they are composed of the same members (representative examples are included in the Supplementary Information). Double-mirror symmetric designs have even more stringent symmetry constraints, making it difficult for even one feature to be present (let alone multiple), inherently limiting the potential for features to interfere with each other within such designs. As such, the impact of a feature on metamaterial properties goes beyond the mere presence of that feature. This impact is likely dependent on the symmetry of a design as well as the presence of other features with different expected influence on properties.  Furthermore, these results highlight the benefits of asymmetry over the other two types of symmetry: asymmetric designs are the most capable of containing multiple features that function as intended without interfering with each other.


\section{Conclusions}
\label{sec:conclusions}

In this study, we examined the mechanical properties of asymmetric, mirror symmetric, and double-mirror symmetric lattice-based metamaterial designs to reveal the expanded options for metamaterial mechanical properties when deviating from double-mirror symmetry. We defined a standardized design space in order to study differences between asymmetric, mirror symmetric, and double-mirror symmetric lattice metamaterial patterns.  We then employed a generative design process to create sets of designs for each type of symmetry.  The composition of these sets are representative of each symmetry-defined design subspace.  The stiffness tensor of each design was then found using a reduced-order finite element model, and the volume fraction of each design was calculated analytically. We quantified the differences among the design spaces using probability distributions of normal stiffness, Poisson's ratio, and normalized shear stiffness.  These distributions indicate that designs of all three types of symmetry are likely to have distinct metamaterial properties from each other. 

Four design features were identified in the design spaces:  \textit{pivot points} and \textit{arrows} were specifically thought to influence Poisson's ratio, while \textit{spider nodes} and \textit{stacked members} were expected to influence shear stiffness.  These features were found to be more common in asymmetric and mirror symmetric designs than in double-mirror-symmetric designs, which was expected as the double-mirror symmetric design space was thought to be less accommodating of these features than the other two types of symmetry.  Probability distributions were then generated to determine the correlation of these features with the metamaterial property space.  All four design features were found to impact metamaterial properties to some degree.  Asymmetric and double-mirror symmetric designs with \textit{arrows} were more likely to have a Poisson's ratio farther from zero than designs without \textit{arrows}.  Similarly, asymmetric and double-mirror symmetric designs with \textit{spider nodes} were more likely to have a larger normalized shear stiffness than designs without \textit{spider nodes}.  However, mirror symmetric designs in both cases were not impacted by these two features in the manner expected. This unexpected result can be explained by examining the presence of multiple types of features in designs.  These results suggest that the impact of a feature on a design is not just tied to the mere presence of that feature: it is also tied to the overlap of multiple features as well as the symmetry type of that design.  Furthermore, asymmetry is shown to be more adept than mirror and double-mirror symmetry at accommodating multiple features that do not inadvertently influence each other's performance.  Because asymmetry is both more versatile and more effective than either mirror or double-mirror symmetry in incorporating design features, it is thus a useful tool for designing lattice patterns to achieve unique metamaterial properties.  As such, this study provides a framework for leveraging asymmetry to identify novel lattice patterns with properties beyond the double-mirror symmetric metamaterial property space.  Future work could examine the impact of a less-discrete design space, as well as expand these methods to 3D lattice patterns, in order to understand how consistent the benefits of asymmetry are with broader design spaces.

\section*{Acknowledgements}
This work is supported and funded by the National Science Foundation under CMMI Nos. 1825444 and 1825521.

\bibliographystyle{elsarticle-harv}
\bibliography{asymm}

\end{document}